\documentclass[final,nofootinbib,aps,pre,twocolumn,showkeys,superscriptaddress,preprintnumbers,floatfix]{revtex4-1}

\usepackage{tabularx}
\usepackage{dynlearn}

\newcommand{\kB}{ k_\text{B} }
\newcommand{\LeftISt}{ L }
\newcommand{\RightISt}{ R }

\colorlet{cmu_color}{black}
\colorlet{cg_color} {blue}
\colorlet{cl_color} {red}
\colorlet{ee_color} {green!66!black}

\begin{document}

\def\ourTitle{%
Harnessing Fluctuations in Thermodynamic Computing\\
via\\
Time-Reversal Symmetries\\
{\small
\vspace{0.1in}
\ourSummary
}
%
}

\def\ourSummary{%
One Sentence Summary:\\
Distinct distributions of thermodynamic work identify signatures of successful
and failed information processing in a microscale flux qubit
}

\def\ourAbstract{%
\bf
We experimentally demonstrate that highly structured distributions of work
emerge during even the simple task of erasing a single bit. These are
signatures of a refined suite of time-reversal symmetries in distinct
functional classes of microscopic trajectories. As a consequence, we introduce
a broad family of conditional fluctuation theorems that the component work
distributions must satisfy. Since they identify entropy production, the
component work distributions encode both the frequency of various mechanisms of
success and failure during computing, as well giving improved estimates of the
total irreversibly-dissipated heat. This new diagnostic tool provides strong
evidence that thermodynamic computing at the nanoscale can be constructively
harnessed. We experimentally verify this functional decomposition and the new
class of fluctuation theorems by measuring  transitions between flux states in
a superconducting circuit.
}

\def\ourKeywords{%
  Jarzynski integral fluctuation theorem, Crooks detailed fluctuation theorem,
  Landauer's Principle, thermodynamics of computation, information
  thermodynamics
}

\hypersetup{
  pdfauthor={James P. Crutchfield},
  pdftitle={\ourTitle},
  pdfsubject={\ourAbstract},
  pdfkeywords={\ourKeywords},
  pdfproducer={},
  pdfcreator={}
}

\author{Gregory Wimsatt}
\email{gwwimsatt@ucdavis.edu}
\affiliation{Complexity Sciences Center and Physics Department, University of California at Davis, One Shields Avenue, Davis, CA 95616}

\author{Olli-Pentti Saira}
\email{osaira@caltech.edu}
\affiliation{Condensed Matter Physics and Kavli Nanoscience Institute,
California Institute of Technology, Pasadena, CA 91125}

\author{Alexander B. Boyd}
\email{abboyd@ucdavis.edu}
\affiliation{Complexity Sciences Center and Physics Department, University of California at Davis, One Shields Avenue, Davis, CA 95616}

\author{Matthew H. Matheny}
\email{matheny@caltech.edu}
\affiliation{Condensed Matter Physics and Kavli Nanoscience Institute,
California Institute of Technology, Pasadena, CA 91125}

\author{Siyuan Han}
\email{han@ku.edu}
\affiliation{Department of Physics and Astronomy, University of Kansas,
Lawrence, KS 66045}

\author{Michael L. Roukes}
\email{roukes@caltech.edu}
\affiliation{Condensed Matter Physics and Kavli Nanoscience Institute,
California Institute of Technology, Pasadena, CA 91125}

\author{James P. Crutchfield}
\email{chaos@ucdavis.edu; Corresponding author}
\affiliation{Complexity Sciences Center and Physics Department, University of California at Davis, One Shields Avenue, Davis, CA 95616}
\affiliation{Condensed Matter Physics and Kavli Nanoscience Institute,
California Institute of Technology, Pasadena, CA 91125}

\date{\today}
\bibliographystyle{unsrt}


\title{\ourTitle}

\begin{abstract}
	\ourAbstract
\end{abstract}

\keywords{\ourKeywords}

\pacs{
  89.75.Kd  
  89.70.+c  
  05.45.Tp  
  02.50.Ey  
  02.50.-r  
  02.50.Ga  
}


\date{\today}
\maketitle

\setstretch{1.1}

\listoffixmes

 

Physics dictates that all computing is subject to spontaneous error. These
days, this truism repeatedly reveals itself: despite the once-predictable
miniaturization of nanoscale electronics, computing performance increases have
dramatically slowed in the last decade or so. In large measure, this is due to
the concomitant rapid decrease in the number of information-bearing physical
degrees of freedom, rendering information storage and processing increasingly
susceptible to corruption by thermal fluctuations. All computing is
thermodynamic. Controlling the production of fluctuations and removing heat
pose key technological challenges to further progress. One comparable setting
that gives some optimism, though, is the overtly functional behavior exhibited
by biological cells---presumably functional information processing by small
numbers of molecules subject to substantial thermal fluctuations. Computing
technologies are very far away from this kind of robust information processing.

Only recently have tools appeared that precisely describe what trade-offs exist
between thermodynamic resources and useful information processing---these are
highly reminiscent of the centuries-old puzzle of how Maxwell's ``very
observant and neat-fingered'' demon uses its ``intelligence'' to convert
disorganized heat energy to useful work \cite{MaxwellToTait1867}. In our modern
era, his demon has led to the realization that information itself is physical
\cite{Land61a,Benn82,Parr15a}---or, most constructively, that information is a
thermodynamic resource \cite{Boyd16d}. This has opened up the new paradigm of
\emph{thermodynamic computing} in which fluctuations play a positive role in
efficient information processing on the nanoscale. We now conceptualize this
via \emph{information engines}: physical systems that are driven by,
manipulate, store, and dissipate energy, but simultaneously generate, store,
lose, communicate, and transform information. In short, information engines
combine traditional engines comprised of heat, work, and other familiar
reservoirs with, what we now call, \emph{information reservoirs}
\cite{Mand012a,Boyd14b}.

Reliable thermodynamic computing requires detecting and controlling
fluctuations in informational and energetic resources and in engine
functioning. For this, one appeals to fluctuation theorems that capture exact
time-reversal symmetries and predict entropy production leading to irreversible
dissipation
\cite{Boch81a,Evan94a,Jarz97a,Croo98a,Croo99a,Seif12a,Klag13a}. We are
now on the door-step of the very far-from-equilibrium thermodynamics needed to
understand the physics of computing. And, in turn, this has started to reveal
the physical principles of how nature processes information in the service of
biological functioning and survival.

Proof-of-concept experimental tests have been carried out in several
substrates: probing biomolecule free energies \cite{Mara08a,Juni09a,Alem12a},
work expended during elementary computing (bit erasure)
\cite{Lamb11a,Beru13a,Mada14a,Jun14a,Beru15a,Hong16a}, and Maxwellian demons
\cite{Kosk15a}.
That said, the suite of contemporary principles
(Supplementary Materials (SM) \ref{app:Principles}) far outstrips experimental
validation to date.


To close the gap, we show how to diagnose thermodynamic computing on the
nanoscale by explaining the signature structures in work distributions
generated during information processing. These structures track the mesoscale
evolution of a system's \emph{informational states} and reveal classes of
functional and nonfunctional microscopic trajectories. We show that the
informational-state evolutions are identified by appropriate conditioning and
that they obey a suite of trajectory-class fluctuation theorems, which
give accurate bounds on work, entropy production, and dissipation. The result
is a new tool that employs mesoscopic measurements to diagnose nanoscale
thermodynamic computing. For simplicity and to make direct contact with
previous efforts, we demonstrate the tools on Landauer erasure of a bit of
information in a superconducting flux qubit.

\begin{figure*}[ht!]
\centering
\includegraphics[width=0.8\linewidth]{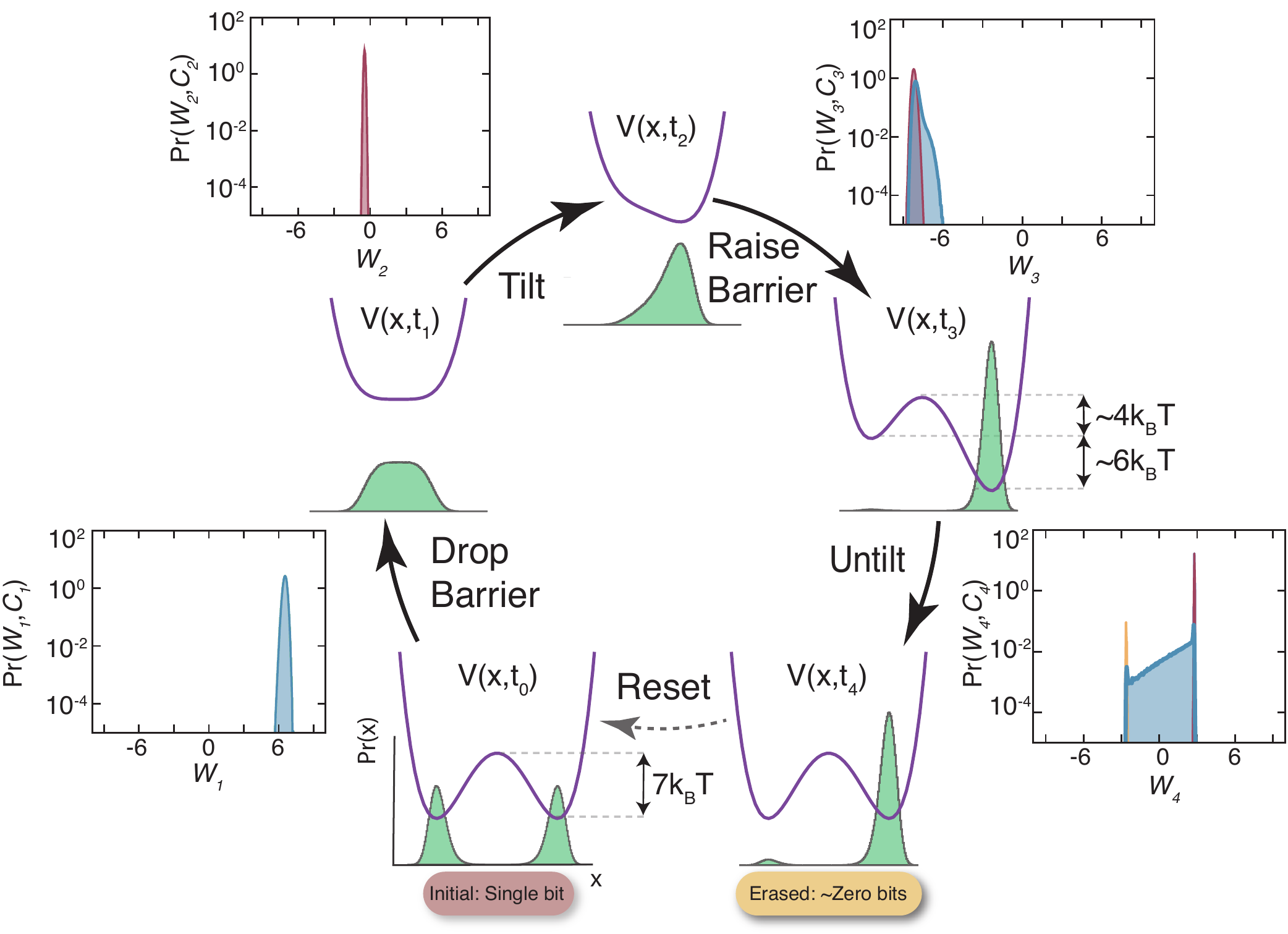}\\
\caption{Inner plot sequence: Erasure protocol (Table \ref{tab:protocol})
	evolution of position distribution $\Pr(x)$. Potential $V(x,t_s)$ at
	substage boundary times $t_s, s = 0, 1, 2, 3, 4$. Starting at $t = t_0$, the
	potential evolves clockwise, ending at $t = t_4$ in the same configuration
	as it starts: $V(x,t_0) = V(x,t_4)$. However, the final position distribution
	$\Pr(x)$ predominantly indicates the $\RightISt$ state.  The original one
	bit of information in the distribution at time $t = t_0$ has been erased.
	Outer plot sequence: Substage work distributions $\Pr(W_s, C_s)$ during
	substages $s$: (1) Barrier Drop, (2) Tilt, (3) Barrier Raise, (4) Untilt.
	During each substage $s$, distributions are given for up to three 
	substage trajectory classes $C_s$: red are of trajectories always in the
	$\RightISt$ state, orange are of trajectories always in the $\LeftISt$
	state, and blue are of the rest, spending some time in each state.
	}
\label{fig:protocol}
\end{figure*}

As a reference, we first explore the thermodynamics of bit erasure in a simple
model: a particle with position and momentum in a double-well potential $V(x,
t)$ and in contact with a heat reservoir at temperature $T$. (Refer to Fig.
\ref{fig:protocol}.) An external controller adds or removes energy from a work
reservoir to change the form of the potential $V(\cdot, t)$ via a predetermined
\emph{erasure protocol} $\{\left(\beta(t),\delta(t)\right): 0 \leq t \leq
\tau\}$. $\beta(t)$ and $\delta(t)$ change one at a time piecewise-linearly
through four protocol substages: (1) \emph{drop barrier}, (2) \emph{tilt}, (3)
\emph{raise barrier}, and (4) \emph{untilt}. (See SM \ref{sec:ThermoWorks}.)
The system starts at time $t=0$ in the equilibrium distribution for a
double-well $V(x, 0)$ at temperature $T$. Being equiprobable, the informational
states associated with each of the two wells thus contain $1$ bit of
information \cite{Cove06a}.

The default potential, $V(\cdot, 0) = V(\cdot, \tau)$, has two symmetric wells
separated by a barrier. Following common practice we call the two wells, from
negative to positive position, the \emph{Left} ($\LeftISt$) and \emph{Right}
($\RightISt$) \emph{informational states}, respectively.

The erasure protocol is designed so that the particle ends in the $\RightISt$
state with high probability, regardless of its initial state. Conducting our
simulation $3.5 \times 10^6$ times, $96.2\%$ of the particles were successfully
erased into the $\RightISt$ state. Thus, as measured by the Shannon entropy,
the initial $1$ bit of information was reduced to $0.231$ bits. Note that we
chose the protocol to give partially inaccurate erasure in order to illustrate
our main results on diagnosing success \emph{and} failure.

At all other times $t$, $V(\cdot, t)$ has either one or two local minima,
naturally defining metastable regions for a particle to be constrained and
gradually evolve towards local equilibrium.  We therefore define the
informational states at time $0 \leq t \leq \tau$ to be the metastable regions,
labeling them $\RightISt$ and, if two exist, $\LeftISt$ --- from most positive
to negative in position.

Since the protocol is composed of four simple substages, we coarse-grain the
system's response by its activity during each substage at the level of its
informational state. Specifically, for each substage, we assign one of three
\emph{substage trajectory classes}: the system (i) was always in the
$\RightISt$ state, (ii) was always in the $\LeftISt$ state, or (iii) spent time
in each. Sometimes there is only one informational state and so the latter two
classes are not achievable for all substages.


We then focus on a single mesoscopic observable---the thermodynamic work
expended during erasure. An individual realization generates a trajectory of
system microstates, with $W(t, t')$ being the work done on the system
between times $0\leq t < t' \leq \tau$; see SM \ref{sec:ThermoWorks}. Let
$W_s = W(t_{s-1}, t_s)$ denote the work generated during substage $s$ and $C_s$
the substage trajectory class. Figure \ref{fig:protocol} (Outer plot sequence)
shows the corresponding substage work distributions $\Pr(W_s, C_s)$ obtained
from our simulations. (See SM \ref{sec:SubstageWorkDists}.)

The drop-barrier and tilt substage work distributions are rather simple, being
narrow and unimodal. The raise-barrier distributions have some asymmetry, but
are also similarly simple. However, the untilt work distributions (farthest
right in Fig. \ref{fig:protocol}) exhibit unusual features that are significant
for understanding the intricacies of erasure. Trajectories that spend all of
the untilt substage in either the $\RightISt$ state or $\LeftISt$ state form
peaks at the most positive (red) and negative (orange) work values,
respectively. This is because the $\RightISt$-state well is always increasing
in potential energy while the $\LeftISt$-state well is always decreasing during
untilt. In contrast, the other trajectories contribute a log-linear ramp of
work values (blue) dependent on the time spent in each. The ramp's positive
slope signifies that more time is typically spent in the $\RightISt$ state.



\begin{figure}[ht!]
\centering
\includegraphics[width=\linewidth]{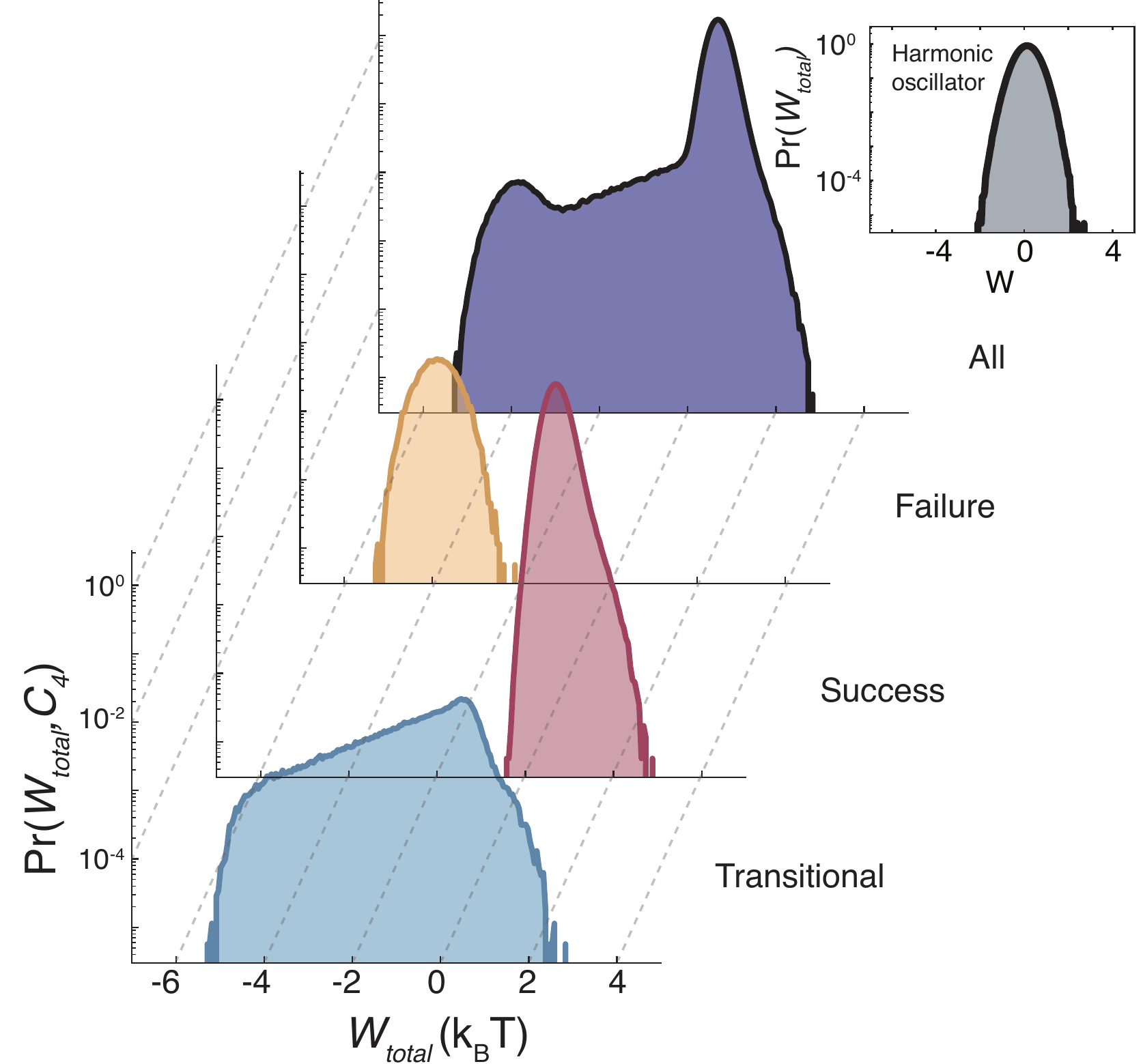}
\caption{
	(Rear, purple) Total work distribution of all trajectories
	$\Pr(W_\text{total})$ during erasure: A histogram generated from
	$3.5 \times 10^6$ trials for $W_\text{total} \in [-6,4]$ over $201$ bins.
	(Inset, gray) Typical unimodal work
	distribution illustrated for spatially-translated thermally-driven simple
	harmonic oscillator.
	(Three front plots) Work distributions $\Pr(W_\text{total}, C_4)$ for the
	trajectory classes $C_4$ determined by the untilt trajectory partition: The
	red work distribution (middle) is that of Success trajectories, the orange
	(rear) is that of Fail trajectories, and the blue (front) is that of the
	remaining, Transitional trajectories.
	}
\label{fig:works}
\end{figure}

Looking at the total work $W_\text{total} = W(0, \tau)$ generated for each
trajectory over the course of the entire erasure protocol, we observe the
strikingly complex and structured distribution $\Pr(W_\text{total})$ shown in
Fig. \ref{fig:works}(Rear). There are two clear peaks at the most positive and
negative work values separated by a ramp.  This highly structured work
distribution, generated by bit erasure, contrasts sharply with the unimodal
work distributions common in previous studies; see, for example, Fig.
\ref{fig:works}(inset) for the work distribution generated by a
thermodynamically-driven simple harmonic oscillator translated in space
or Fig. 2 in Ref. \cite{Croo99a}.

We can understand the mechanisms behind this structure when decomposing Fig.
\ref{fig:works} (Rear)'s total work distribution under the untilt substage
trajectory classes $C_4$. We label trajectories that spend all of the untilting
substage in the $\RightISt$ state \emph{Success} since, via the previous
substages, they reach the intended
$\RightISt$ state by the untilting substage and remain there until the
protocol's end. Similarly, trajectories that spend all of the untilt substage
in the $\LeftISt$ state are labeled \emph{Fail}. The remaining trajectories are
labeled \emph{Transitional}, since they transition between the two
informational states during untilt, potentially succeeding or failing to end in
the $\RightISt$ state. Figure \ref{fig:works} (Three front plots) shows the
work distribution for each of these three trajectory classes. Together they
recover the total work distribution over all trajectories shown in Fig.
\ref{fig:works}(Rear). Though, now the thermodynamic contributions to the total
from the functionally-distinct component trajectories are made apparent.


\newcommand{\alltraj}{{\overrightarrow {\mathcal{Z}}}}
\newcommand{\traj}{{\overrightarrow z}}
\newcommand{\Rev}{\texttt{R}}
\newcommand{\rev}[1]{#1^\texttt{R}}
\newcommand{\Feq}{F}
\newcommand{\forP}{\mathcal{P}}
\newcommand{\revPeq}{\mathcal{R}}
\newcommand{\partition}{Q}



Exploring the mesoscale dynamics of erasure revealed signatures of a
``thermodynamics'' for each trajectory that is closely associated with
successful or failed information processing. We now introduce the underlying
fluctuation theory from which the trajectory thermodynamics follow. Key
to this is comparing system behaviors in both forward and reverse time
\cite{Boch81a,Evan94a,Jarz97a,Croo98a,Croo99a,Seif12a,Klag13a}. (See SM
\ref{app:TCFTs} and \ref{app:tcft_derivations}.)

This suite of trajectory-class fluctuation theorems (TCFTs) applies to
arbitrary classes of system microstate trajectories obtainable during a
thermodynamic transformation. Importantly, they interpolate between Jarzynki's
equality \cite{Jarz97a} and Crooks' detailed fluctuation theorem
\cite{Croo99a}, as the trajectory class varies.
This lower bounds the average work $\langle W \rangle_C$ over any measurable
subset $C$ of the ensemble of system microstate trajectories $\alltraj$, where
$W$ is the total work for a trajectory:
\begin{align}
\langle W \rangle_C
&\geq \Delta F + \kB T \ln \frac{\forP(C)}{\revPeq(\rev C)} \nonumber \\
&= \langle W \rangle_C^\text{min}
\label{eq:tcft_avgwork_bound}
 ~,
\end{align}
with $\Delta \Feq$ the change in equilibrium free energy over the protocol,
$\forP(C)$ the probability of realizing the class $C$ during the protocol, and
$\revPeq(\rev C)$ the probability of obtaining the time reverse of class $C$
under the time-reverse protocol. $\kB$ is Boltzmann's constant.

The TCFTs lead to several consequences. First, well-formulated trajectory
classes allow accurate estimates of the works for their trajectories, even with
limited knowledge of system response under the protocol and its time reverse.
Second, they strictly and more strongly bound the average work over all
trajectories compared to the equilibrium free energy change $\Delta \Feq$.
Third, they provide a new expression for obtaining equilibrium free-energy
changes:
\begin{align}
\label{eq:tcft_deltaFeq}
\Delta \Feq = -\kB T \ln \left( \frac{\forP(C)}{\revPeq(\rev C)}
\langle e^{-W / \kB T} \rangle_C \right)
 ~.
\end{align}
Remarkably, this only requires statistics for a particular class $C$ and its
reverse $\rev C$ to produce the system's free energy change. Since rare
microstate trajectories may generate sufficiently negative works that dominate
the average exponential work, this leads to a substantial statistical advantage
over direct use of Jaryznski's equality $\Delta \Feq = -\kB T \ln \langle
e^{-W / \kB T} \rangle_\alltraj$ for estimating free energies \cite{Jarz06a}.


\begin{figure*}[ht!]
\centering
\includegraphics[width=0.9\linewidth]{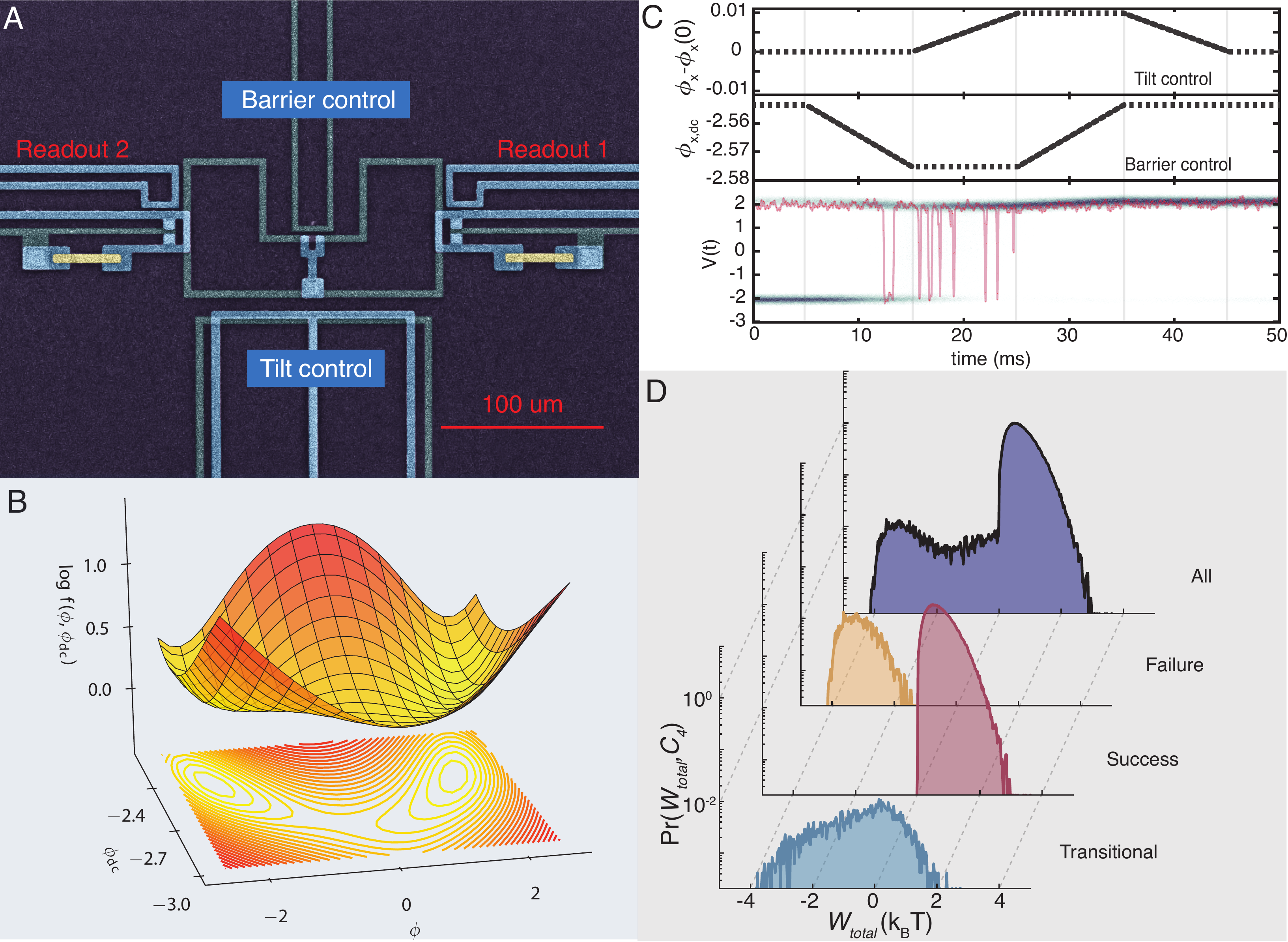}
\caption{Superconducting implementation of metastable memory and bit erasure
	driven by thermal fluctuations:
	(A) Optical micrograph of a gradiometric flux qubit with control lines and
	local magnetometers for state readout. The flux $\phi_\text{x}$, threading
	the large U-shaped differential-mode loop, controls the potential's tilt and
	flux $\phi_\text{xdc}$, threading the small SQUID loop, controls the
	potential barrier height. Currents in the \emph{barrier control} and
	\emph{tilt control} lines modulate those fluxes.
	(B) Calculated potential energy landscape at the beginning of the erasure
	protocol; see Eqs. (\ref{eq:FluxQubitEoM}) and
	(\ref{eq:FluxQubitPotential}).
	(C top) Sequence of tilt and barrier control waveforms implementing bit
	erasure and (C bottom) sample of resulting magnetometer traces tracking the
	system's internal state.
	(D) Work distributions $\Pr(W_\text{total}|C_4)$ over trajectories
	conditioning on the Success, Fail, and Transitional classes.
	Experimental distributions obtained from $10^5$ protocol repetitions.
	}
\label{fig:FluxQubit}
\end{figure*}


To explore these predictions, we selected a superconducting flux qubit composed
of paired Josephson junctions (Fig. \ref{fig:FluxQubit}(A)), resulting in a
double-well nonlinear potential that supports information storage and
processing (Fig. \ref{fig:FluxQubit}(B)). SM \ref{sec:FluxQubitPhysics}
explains the physics underlying their nonlinear equations of motion, comparing
the similarities and differences with our model's idealized Langevin dynamics.

Despite control protocols for double-well potentials that perform accurate and
efficient bit erasure \cite{Boyd18a}, we run the flux qubit in a mode that
yields imperfect erasure (Fig. \ref{fig:FluxQubit}(C)). As with the
simulations, our intention is to illustrate how trajectory classes and the
TCFTs can be used to diagnose and interpret success and failure in microscopic
information processing using only mesoscopic measurements of work.

Interplay between the geometric, linear magnetic inductances and the nonlinear
Josephson inductances gives rise to a potential landscape that can be
controlled with external bias fluxes. It is natural to call the $\phi_\text{x}$
and $\phi_\text{xdc}$ fluxes, threading the differential mode and the small
SQUID loop, respectively, the \emph{tilt} and \emph{barrier controls}. (See
(Fig. \ref{fig:FluxQubit}(A) caption.) SM \ref{sec:FluxQubit} presents a
derivation of the flux qubit potential and details its calibration. All
experiments presented here were carried out at a temperature of $500$ mK.

To execute an erasure protocol, we first choose an information-storage state
with a tall barrier and two equal-depth wells. The two-dimensional potential
for this at the calibrated device parameters is depicted in Fig.
\ref{fig:FluxQubit}(B). We implement the bit erasure protocol as a time-domain
deformation imposed by the two control fluxes that starts and ends at the
storage configuration. The amplitudes of the control waveforms in reduced units
are small; see Fig. \ref{fig:FluxQubit}(C). Hence, the microscopic energetics
change linearly as a function of the control fluxes.

We use a local dc-SQUID magnetometer to continuously monitor the trapped flux
state in the device---\emph{Readout} 1 in Fig. \ref{fig:FluxQubit}(A). The
digitized signal has a rise time of $100~\mu$s, after which the two logical
states are discriminated virtually without error. A typical magnetometer trace
$V(t)$ acquired during the execution of the erasure protocol is shown in Fig.
\ref{fig:FluxQubit}(C). We operate the magnetometer with a low-amplitude AC
current bias at 10 MHz to avoid an increase in the effective temperature during
continuous readout of the flux state due to wideband electromagnetic
interference.

To collect work statistics, we repeat the erasure protocol $10^5$ times. We
identify the logical-state transitions from the magnetometer traces as
zero-crossings, recording the direction $\delta_i$---sign convention: +1 ($-$1)
for a L-to-R (R-to-L) transition---and the time $t_i$ relative to the start
of the protocol. We evaluate a single-shot work estimate $W = \sum_i \delta_i
U_{LR}(t_i)$, where $U_{LR}(t) = U_R(t) - U_L(t)$ is the biasing of
the potential minima at time $t_i$. Making use of the linearity of the
system energetics and the choice of offsets and compensation coefficients, we
find $U_{LR}(t) = A \left(\phi_x(t) - \phi_x(0) \right)$, with the
coefficient $A = 210 \text{K} \times \kB$ evaluated from the calibrated potential. The
above work estimate based on the logical-state transitions is an accurate
estimate of the true microscopic work assuming that the timescales for the
state transitions and for changes in the control parameters are much slower
than the intra-well equilibration. (See SM \ref{sec:ThermoWorks}.)

The total work distribution estimated from the flux qubit experiments is shown
as the rear-most distribution in Fig. \ref{fig:FluxQubit}(D). Using the
previous microstate trajectory partitioning into the \emph{Success,}
\emph{Fail,} and \emph{Transitional} trajectory classes reveals a decomposition
of the total work distribution given by Fig. \ref{fig:FluxQubit}(D)(Three front
panels). The close similarity with our simulations (Fig. \ref{fig:works}) is
notable. Especially so, given the rather substantial differences between the
simulated system (idealized double-well potential and thermal noise, exactly
one-dimension system, ...) and the experimental system (complex potential in
two dimensions, nonideal fluctuations, ...). \emph{A priori} it is not clear
that the theoretical predictions of the informational classes should apply so
directly and immediately to the real-world qubit. In point of fact, these
differences serve to emphasize the descriptive power of the mesoscopic work
fluctuation theorems: despite substantial differences in system detail they
successfully diagnose the information-processing classes of microscopic
trajectories. This robustness will be especially helpful in monitoring
thermodynamic computing in biological systems, where, in many cases,
information-bearing degrees-of-freedom cannot be precisely modeled.



We experimentally demonstrated that work fluctuations generated by information
engines are highly structured. Nonetheless, they strictly obeyed a suite of
time-reversal symmetries---the trajectory-class fluctuation theorems introduced
here. The latter are direct signatures of how a system's informational
states evolve and they identify functional and nonfunctional microscopic
trajectory bundles. We showed that the trajectory-class fluctuation theorems
naturally interpolate between Jarzynski's integral and Crooks' detailed
fluctuation theorems, providing a unified diagnostic probe of nonequilibrium
thermodynamic transformations that support information processing.

Using them we gave a detailed mechanistic analysis of the thermodynamics of the
now-common example of erasing a bit of information as an external protocol
manipulated a stochastic particle in a double-well potential (simulation) and
the stochastic state of a flux qubit (experiment). To give insight into the new
level of mechanistic analysis possible, we briefly discussed the untilt
trajectory-class partitioning. Though ignoring other protocol stages, this was
sufficient to capture the basic trajectory classes that generate the overall
work distribution's features. Partitioning on informational-state occupation
times during barrier raising and untilting---an alternative used in follow-on
studies---yields an even more incisive decomposition of the work distributions
and diagnosis of informational functioning. The corresponding bounds on
thermodynamic resources obtained via the TCFTs also improve on current
estimation methods. The net result is that trajectory-class analysis can be
readily applied to debug thermodynamic computing by engineered or biological
systems.


\bibliography{chaos,ref,greg}

\paragraph*{Acknowledgments:}
\label{sec:acknowledgments}

We thank C. Jarzynski, D. Mandal, and P. Riechers for helpful discussions. As
an External Faculty member, JPC thanks the Santa Fe Institute and the Telluride
Science Research Center for their hospitality during visits.

\paragraph*{Funding:}
\label{sec:Funding}
This material is based upon work supported by, or in part by, the U. S. Army
Research Laboratory and the U. S. Army Research Office under contracts
W911NF-13-1-0390 and W911NF-18-1-0028.

\paragraph*{Author Contributions Statement:}
\label{sec:Contrib}
JPC, OPS, MLR, and GW conceived of the project. ABB, GW, and JPC developed the
theory. SH provided the flux qubit and the experimental and analytical methods
for its calibration. MHM, MLR, OPS, designed, implemented, and carried out the
experiments. GW and JPC performed the calculations. JPC, OPS, MLR, and GW
drafted the manuscript. JPC and MLR supervised the project.

\paragraph*{Competing Financial Interests Statement:}
\label{sec:COI}
The authors declare that they have no competing financial interests.

\paragraph*{Data Availability:}
The data that support the findings of this study are available from the corresponding author on reasonable request.

\subsection*{Supplementary Materials}
Materials and Methods: Derivation of trajectory-class
fluctuation theorems, further discussion and interpretation, and experiment
implementation, calibration, and work estimation methods.

\onecolumngrid
\newpage
\begin{center}
\large{Supplementary Materials}\\
\vspace{0.1in}
\emph{\ourTitle}\\
\vspace{0.1in}
{\small
Gregory Wimsatt, Olli-Pentti Saira, Alexander B. Boyd, Matthew H. Matheny,\\
Siyuan Han, Michael L. Roukes, and James P. Crutchfield
}
\end{center}

\setcounter{equation}{0}
\setcounter{figure}{0}
\setcounter{table}{0}
\setcounter{page}{1}
\makeatletter
\renewcommand{\theequation}{S\arabic{equation}}
\renewcommand{\thefigure}{S\arabic{figure}}
\renewcommand{\thetable}{S\arabic{table}}

%

\begin{center}
\large
{\bf Materials and Methods}
\normalsize
\end{center}

The following presents details, derivations, and explanations for the
theoretical claims and experimental results. First, we give a synopsis of
recently developed principles of thermodynamic computing. Then, we introduce
the model microscopic stochastic thermodynamical system, including its
equations of motion and its physical calibration. The following sections
explain how to use and interpret the trajectory-class fluctuation theorems and
provide their derivation. Their practical application to work estimation for
trajectories and classes, in light of the alternative kinds of work (inclusive
and exclusive) and methods for experimental estimation, is laid out. A brief
commentary on the substage work distributions then follows. Finally, we turn to
describe the flux qubit, its equations of motion, implementation, calibration,
and measurement.

\section{Principles of Thermodynamic Computing: A recent synopsis}
\label{app:Principles}

A number of closely-related thermodynamic costs of computing have
been identified, above and beyond the \emph{house-keeping heat} that maintains
a system's overall nonequilibrium dynamical state. First, there is the
\emph{information-processing Second Law}
\cite{Boyd15a}
that extends
Landauer's original bound on erasure
\cite{Land61a}
to dissipation in general computing
and properly highlights the
central role of information generation measured via the physical substrate's
dynamical Kolmogorov-Sinai entropy.
It specifies the minimum amount of energy
that must be supplied to drive a given amount of computation forward. Second,
when coupling thermodynamic systems together, even a single system and a
complex environment, there are transient costs as the system synchronizes to,
predicts, and then adapts to errors in its environment
\cite{Stil12a,Boyd16e,Boyd16c}.
Third, the very modularity of a
system's organization imposes thermodynamic costs
\cite{Boyd17a}.
Fourth, since
computing is necessarily far out of equilibrium and nonsteady state, there are
costs due to driving transitions between information-storage states
\cite{Riec16b}.
Fifth, there are costs to generating randomness
\cite{Agha16d},
which is itself a widely useful resource. Finally, by way of harnessing these
principles, new strategies for optimally controlling nonequilibrium
transformations have been introduced
\cite{Zulk15a,Ging16a,Patr17a,Boyd18a}.

\section{Microscopic Stochastic Thermodynamical System}
\label{supp:Model}

For concreteness, we concentrate on a one-dimensional system: a particle with
position and momentum in an external potential $V(x, t)$ and in contact with a
heat reservoir at temperature $T$. An external controller adds or removes
energy from a work reservoir to change the form of the potential $V(\cdot, t)$
via a predetermined \emph{erasure protocol} $\{\left(\beta(t),\delta(t)\right):
0 \leq t \leq \tau\}$. (See Supplementary Materials (SM) \ref{sec:ThermoWorks}
for details on the alternative definitions of work.) The potential takes the
form:
\begin{align*}
V(x, t) = a x^4 - b_0 \beta(t) x^2 - c_0 \delta(t) x
~,
\end{align*}
with constants $a, b_0, c_0 > 0$. During the erasure protocol, $\beta(t)$ and
$\delta(t)$ change one at a time piecewise-linearly through four protocol
substages: (1) \emph{drop barrier}, (2) \emph{tilt}, (3) \emph{raise barrier},
and (4) \emph{untilt}, as shown in Table \ref{tab:protocol}. The system starts
at time $t=0$ in the equilibrium distribution for a double-well $V(x, 0)$ at
temperature $T$. Being equiprobable, the informational states associated with
each of the two wells thus contain $1$ bit of information \cite{Cove06a}.
The effect of the control protocol on the system potential and system response
is graphically displayed in Fig. \ref{fig:protocol}.

\begin{table}[ht]
\begin{center}
{\setlength{\extrarowheight}{3pt}%
\begin{tabular}{|c|| c @{\hspace{-1pt}} c @{\hspace{-1pt}} c @{\hspace{-1pt}} c
@{\hspace{-1pt}}  c @{\hspace{-1pt}} c @{\hspace{-1pt}} c @{\hspace{-1pt}} c
@{\hspace{-1pt}} c|}
\hline
Stage &&Drop Barrier&&Tilt&&Raise Barrier&&Untilt& \\
$t_s$  & $t_0$ && $t_1$ && $t_2$ && $t_3$ && $t_4$ \\
\hline
$\beta(t)$ &$\biggr\vert$& $\frac{t_1-t}{t_1-t_0}$ &$\biggr\vert$& 0 &$\biggr\vert$&
$\frac{t-t_2}{t_3-t_2}$ &$\biggr\vert$& $1$ &$\biggr\vert$ \\
$\delta(t)$ & $\biggr\vert$& 0 &$\biggr\vert$& $\frac{t-t_1}{t_2-t_1}$ &$\biggr\vert$& $1$ &$\biggr\vert$&
$\frac{t_4 - t}{t_4 - t_3} $ &$\biggr\vert$ \\
\hline
    \end{tabular}}
  \end{center}
\caption{Erasure protocol.
  }
\label{tab:protocol}
\end{table}

We model the erasure physical information processing with underdamped Langevin
dynamics:
\begin{align}
dx & = v dt \nonumber \\
dv & = \sqrt{2 \kB T\gamma/m}\, r(t) \sqrt{dt}
  - \left( \frac{\partial}{\partial x} V(x, t) + \gamma v \right) dt
  ~,
\label{eq:LangevinEoM}
\end{align}
where $\kB$ is Boltzmann's constant, $\gamma$ is the coupling between the heat
reservoir and system, $m$ is the particle's mass, and $r(t)$ is a memoryless
Gaussian random variable with $\langle r(t) \rangle = 0$ and $\langle r(t)
r(t') \rangle = \delta(t-t')$.

For comparison to experiment, we simulated erasure with the following
parameters, sufficient to fully specify the dynamics: $\gamma \tau / m = 500$,
$2 \kB T \tau^2 a / (m b_0) = 2.5 \times 10^5$, $b_0^2 / (4 a \kB T) = 7 $, and
$\sqrt{8a/b_0^3} c_0 = 0.4$.
The resulting potential, snapshotted at times during the erasure substages, is
shown in Fig. \ref{fig:protocol}(Inner plot sequence).

Reliable information processing dictates that we set time scales so that the
system temporarily, but stably, stores information. To support
metastable-quasistatic behavior at all times the relaxation rates of the
informational states are much faster than the rate of change of the potential,
keeping the system near metastable equilibrium throughout. The entropy
production for such protocols tends to be minimized.

%

\section{Trajectory-Class Fluctuation Theorems: Use and Interpretation}
\label{app:TCFTs}

Here, we describe the trajectory-class fluctuation theorems, explaining several
of their possible implications and exploring their application to both the
simulations and flux qubit experiment. Their derivations are given in the
section following.

First, consider a forward process distribution $\forP$, defined by the
probabilities of the system microstate trajectories $\alltraj$ due to an
initial equilibrium microstate distribution evolving forward in time under a
control protocol. Then, the reverse process distribution $\revPeq$ is
determined by preparing the system in equilibrium in the final protocol
configuration and running the reverse protocol. The reverse protocol is the
original protocol conducted in reverse order but also with objects that are odd
under time reversal, like magnetic fields, negated. The time-reversal of a
trajectory $\traj = (z_0, z_1, \ldots z_\ell)$ is $\rev \traj = (\rev{z_\ell},
\ldots, \rev{z_1}, \rev{z_0})$, where $\rev{z_i} = -z_i$ if $z_i$ is odd under
time-reversal (e.g., momentum or spin), otherwise $\rev{z_i} = z_i$. For a
measurable subset of trajectories $C \subset \alltraj$, let $\langle \cdot
\rangle_C$ denote an average over the ensemble of forward process trajectories
conditioned on the trajectory class $C$. Let $\forP(C)$ and $\revPeq(\rev C)$
denote the probabilities of observing the class $C$ in the forward process and
the reverse class $\rev C = \{\rev \traj | \traj \in C\}$ in the reverse
process, respectively.


We first introduce a \emph{trajectory-class fluctuation theorem} (TCFT) for the
class-averaged exponential work $\langle e^{-W / \kB T} \rangle_C$:
\begin{align}
\langle e^{-W / \kB T} \rangle_C
= \frac{\revPeq(\rev C)}{\forP (C)} e^{-\Delta \Feq / \kB T} 
 ~,
\label{eq:tcft_expavg_work}
\end{align}
with $\Delta \Feq$ the system equilibrium free energy change. We also introduce
a \emph{class-averaged work} TCFT:
\begin{align}
\label{eq:tcft_avgwork_equality}
\langle W \rangle_C & = \Delta \Feq + \kB T
  \bigg(
D_\text{KL} \left[ \forP (\alltraj | C) || \revPeq(\rev \alltraj | \rev C) \right]
+ \ln \frac{\forP (C)}{\revPeq(\rev C)}
  \bigg)
 ~.
\end{align}
This employs the Kullback-Liebler divergence $D_\text{KL}[~\cdot~]$ taken
between forward and reverse process distributions over all class trajectories
$\traj \in C$, conditioned on the forward class $C$ and reverse class $\rev C$,
respectively. If we disregard this divergence, which is nonnegative and would
generally be difficult to obtain experimentally, we then find the lower bound
$\langle W \rangle_C^\text{min}$ on the class-averaged work of Eq.
(\ref{eq:tcft_avgwork_bound}).

In the limit of class $C$ possessing only a single trajectory, we recover
detailed fluctuation theorems as in Ref. \cite{Croo99a}. If, however, we take
$C$ to be the entire set of trajectories $\alltraj$, we recover integral
fluctuation theorems as in Jarzynski's equality \cite{Jarz97a}. Thus, the TCFTs
are a suite that spans the space of fluctuation theorems between the extreme of
the detailed theorems, that require very precise information about an
individual trajectory, and the integral theorems, that describe the system's
entire trajectory ensemble. SM \ref{app:tcft_derivations} below provides proofs
for both TCFTs.

We can rearrange Eq. (\ref{eq:tcft_expavg_work}) to obtain Eq.
(\ref{eq:tcft_deltaFeq})---an expression for estimating equilibrium free energy
changes:
\begin{align}
\Delta \Feq = -\kB T \ln \left( \frac{\forP(C)}{\revPeq(\rev C)} \langle
e^{-W / \kB T} \rangle_C \right)
 ~.
\end{align}
Thus, to estimate free energy one sees that statistics are needed for only one
particular class and its reverse. Generally, this gives a substantial
statistical advantage over direct use of Jaryznski's equality:
\begin{align*}
\Delta \Feq = -\kB T \ln \langle e^{-W / \kB T} \rangle _\alltraj
 ~,
\end{align*}
since rare microstate trajectories may generate negative work values that
dominate the average exponential work
\cite{Jarz06a}.
The problem is clear in the case of erasure. Recall from Fig.
\ref{fig:works}(Three front panels) that Fail trajectories generate the
most-negative work values. In the limit of higher success-rate protocols that
maintain low entropy production, failures generate more and more negative
works, leading them to dominate when estimating average exponential works.

In contrast, to efficiently determine the change in equilibrium free energy
from Eq. (\ref{eq:tcft_deltaFeq}), its form indicates that one should choose a
class that (i) is common in the forward process, (ii) has a reverse class that
is common in the reverse process, and (iii) generates a narrow work
distribution. This maximizes the accuracy of statistical estimates for the
three factors on the RHS. For example, while the equilibrium free energy change
in the case of our erasure protocol is theoretically simple (zero); the Success
class fits the criteria.

We can then monitor the class-averaged work in excess of its bound:
\begin{align*}
E_C & = \langle W \rangle_C - \langle W \rangle_C^\text{min} \\
    & = k_B T D_\text{KL} \left[
	\forP ( \alltraj | C) || \revPeq (\rev \alltraj | \rev C)
	\right ] \\
    & \geq 0
 ~.
\end{align*}
The inequality in Eq. (\ref{eq:tcft_avgwork_bound}) is a refinement of the
equilibrium Second Law and therefore the bound $\langle W \rangle_C^\text{min}$
generally provides a more accurate estimate of the average work of trajectories
in a class compared to the equilibrium free energy change $\Delta \Feq$. More
precisely, as we will see below, an average of the excess $E_C$ over all
classes $C$ in a partition of trajectories must be smaller than the dissipated
work $\langle W \rangle - \Delta \Feq$. For trajectory classes with narrow work
distributions, this can be a significant improvement. We can see this by Taylor
expanding the LHS of Eq. (\ref{eq:tcft_expavg_work}) about the mean
dimensionless work $\langle W / \kB T \rangle_C$. This shows that Eq.
(\ref{eq:tcft_avgwork_bound}) becomes an equality when the variance and higher
moments vanish. SM \ref{sec:app_expansion} below delves more into moment
approximations. In any case, trajectory classes with narrow work distributions
have small excess works $E_C$.

To estimate $\revPeq(\rev C)$, we ran $3.5 \times 10^6$ simulations of the
reverse process. Table \ref{tab:tcft} shows that the Success and Fail classes
have small excesses and, as seen in Fig. \ref{fig:works}(Three front panels),
these classes indeed have narrow work distributions. Elsewhere we explore these
and additional
partition schemes, finding that the Transitional trajectories can be further
partitioned to yield narrow work distributions so that all trajectory classes
have small excesses $E_C$. In short, this demonstrates how well-formulated
trajectory classes allow accurate estimates on the works for all trajectories.

\begin{table}[ht]
\begin{center}
{\setlength{\extrarowheight}{3pt}%
\vspace{10pt}

Simulation

\vspace{10pt}
\begin{tabular}{| l || c | c | c |} 
\hline
Class $C$ & $\langle W \rangle_C$ & $\langle W \rangle_C^\text{min}$ & $E_C$ \\
\hline
All $\alltraj$ & 0.634 & 0.0 & 0.634 \\
Success & 0.713 & 0.683 & 0.030 \\
Fail & -3.885 & -3.951 & 0.066 \\
Transitional & -0.546 & -1.650  & 1.170 \\
\hline
    \end{tabular}
\begin{tabular}{| l || c | c | c |}
\hline
Partition $\partition$ & $\langle W \rangle_{\alltraj}$
& $\langle W \rangle_Q^\text{min}$ & $E_Q$ \\
\hline
Trivial $\left\{\alltraj\right\}$ & 0.634 & 0.0 & 0.634 \\
Untilt-Centric I & 0.634 & 0.560 & 0.074 \\
Untilt-Centric II & 0.634 & 0.601 & 0.032 \\
\hline
    \end{tabular}
\vspace{10pt}

Experiment

\vspace{10pt}
\begin{tabular}{| l || c | c | c |} 
\hline
Class $C$ & $\langle W \rangle_C$ & $\langle W \rangle_C^\text{min}$ & $E_C$ \\
\hline
All $\alltraj$ & 0.668 & 0.0 & 0.668 \\
Success & 0.742 & 0.643 & 0.099 \\
Fail & -3.132 & -3.475 & 0.343 \\
Transitional & -0.443 & -1.215  & 0.772 \\
\hline
    \end{tabular}
\begin{tabular}{| l || c | c | c |}
\hline
Partition $\partition$ & $\langle W \rangle_{\alltraj}$
& $\langle W \rangle_Q^\text{min}$ & $E_Q$ \\
\hline
Trivial $\left\{\alltraj\right\}$ & 0.668 & 0.0 & 0.668 \\
Untilt-Centric I & 0.668 & 0.554 & 0.115 \\
\hline
    \end{tabular}}
  \end{center}
\caption{(Top Left) Comparison of simulated class-average works and bounds for
	different trajectory classes: \emph{All} trajectories $\protect\alltraj$,
	\emph{Success} trajectories, \emph{Fail} trajectories, and
	\emph{Transitional} trajectories. These are identified in Fig.
	\ref{fig:works} (Four front panels). From left to right, columns give the
	estimated class-average work $\langle W \rangle_C$, TCFT lower bound
	$\langle W \rangle_C^\text{min}$, and their difference $E_C$. $3.5 \times
	10^6$ simulations were run for each of the forward and reverse processes,
	with $96.2\%$ trajectories successfully ending in the $\RightISt$
	informational state under the forward process.
	(Top Right) Comparison of ensemble-average work and bounds due to different
	partitions: \emph{Trivial} partition; \emph{Untilt-Centric I} partition,
	composed of Success, Fail, and Transitional; and \emph{Untilt-Centric II}
	partition, described in follow-on work. From left to right, columns give
	the estimated ensemble-average work, the partition bound $\langle W
	\rangle_Q^\text{min}$, and their difference $E_Q$. All values in units of
	$\kB T$.
	(Bottom) Parallel results from the flux qubit experiment. \emph{All}
	trajectories $\protect\alltraj$, \emph{Success}, \emph{Fail}, and
	\emph{Transitional} trajectories identified in Fig. \ref{fig:FluxQubit}(D).
	Data from $195,050$ trajectories from the forward protocol and $250,000$
	trajectories from the reverse protocol.
  }
\label{tab:tcft}
\end{table}

To measure the efficacy of a given partition $\partition$ of trajectories into
classes, we ask what the ensemble-average of class-average excess works is:
\begin{align*}
E_\partition & = \sum_{C \in \partition} \forP(C) E_C \\
  &= \langle W \rangle_\alltraj
  - \sum_{C \in \partition} \forP(C) \langle W \rangle_C^\text{min} \\
  &= \langle W \rangle_\alltraj
  - \langle W \rangle_\partition^\text{min}
 ~,
\end{align*}
with
$\langle W \rangle_\partition^\text{min} = \sum_{C \in \partition} \forP(C)
\langle W \rangle_C^\text{min}$.

From Eq. (\ref{eq:tcft_avgwork_bound}), we see that $\langle W
\rangle_\partition^\text{min}$ is the coarse-grained lower bound on
ensemble-average dissipation from Ref. \cite{Gome08a}:
\begin{align*}
\langle W \rangle_\partition^\text{min}
  & = \Delta \Feq + \kB T
  D_\text{KL} \left [ \forP(\partition) || \revPeq(\rev \partition) \right]
  ~,
\end{align*}
where $D_\text{KL} \left[ ~\cdot~ \right]$ is the Kullback-Liebler divergence
between forward and reverse process distributions over the trajectory classes
$C \in \partition$. Since Kullback-Liebler divergences are nonnegative, such a
bound always provides an improvement over the equilibrium Second Law. Table
\ref{tab:tcft} shows both $\langle W \rangle_\partition^\text{min}$ and
$E_\partition$ for the trivial partition $\{ \alltraj \}$, our three-class
partition, labeled \emph{Untilt-Centric I}, and the improved partition
described in follow-on work, labeled \emph{Untilt-Centric II}. In this case,
the latter two also provide an improvement on the nonequilibrium Second Law
which, assuming metastable starting and ending distributions, provides a lower
bound on the average work equal to $0.533$, the change in nonequilibrium free
energy.

We can appeal to Landauer's erasure bound---$k_B T \ln 2 \approx 0.693\, k_B
T$---to calibrate the excesses $E_C$ and $E_Q$. We see for the simulation data
that our three-class partition Untilt-Centric I provides class-average work
bounds that, on average, are only about $11\%$ of $k_B T \ln 2$ from the actual
class-average works. The more refined Untilt-Centric II partition reduces this
excess to about $5 \%$ while the trivial partition fails by about $91 \%$ of
$k_B T \ln 2$.

The experimental data matches these results, with the largest discrepancy
occurring for the class-average excess for the Fail class. This is not wholly
coincidence, since we determined the parameters of the experimental protocol
by adjusting the parameters of a simple two-state Markov simulation to obtain a
work distribution similar to that obtained by our Langevin simulations of the
Duffing potential system described in the main text. However, it is interesting
that this was sufficient to provide matches in both the clean decomposition of
the total work distribution by trajectory class and the quantities of Table
\ref{tab:tcft}.

We also recover the equality of Ref. \cite{Gome08a} for the ensemble-average
work by averaging Eq. (\ref{eq:tcft_avgwork_equality}) over each class:
\begin{align*}
\langle W \rangle &= \sum_C \forP(C) \langle W \rangle_C \\
&= \Delta \Feq + \kB T \bigg(
\sum_{C \in \partition} \forP(C)
  D_\text{KL} \left[
  \forP (\alltraj | C) || \revPeq(\rev \alltraj | \rev C)
  \right]
  + D_\text{KL} \left[
  \forP(\partition) || \revPeq(\rev \partition)
  \right]  \bigg)
 ~,
\end{align*}
which of course is lower bounded by $\langle W \rangle_\partition^\text{min}$.

These results suggest the criterion for optimal trajectory partitions: Select a
partition sufficiently refined to yield tight bounds on class-average works,
but no finer. Machine learning methods for model order selection will provide a
basis for a natural classification scheme for trajectories that captures all
relevant thermodynamics and information processing.

By changing our forward and reverse processes $\forP$ and $\revPeq$ to begin in
system microstate distributions other than equilibrium, a yet-broader class of
TCFTs emerge. We can then find analogous results for heats and comparisons with
works and nonequilibrium free-energy changes. We explore these in depth
elsewhere.

\section{TCFT Derivations}
\label{app:tcft_derivations}

\newcommand{\microst}{{z}}
\newcommand{\microsts}{{\mathcal Z}}

We now present derivations for the two TCFTs introduced in Eqs.
(\ref{eq:tcft_expavg_work}) and (\ref{eq:tcft_avgwork_equality}).

Assume that the system dynamics is described by a Hamiltonian specified in part
by an external control protocol, as well as by a weak coupling to a thermal
environment that induces steady relaxation to canonical equilibrium.

Start the system in equilibrium distribution $\pi_0$ for Hamiltonian $\mathcal
H_0$ and run a protocol until time $\tau$, causing the system Hamiltonian to
evolve to $\mathcal H_\tau$. If we then hold the Hamiltonian at $\mathcal
H_\tau$ for a long time, the system relaxes into the equilibrium distribution
$\pi_\tau$. The system's ensemble entropy change from $t=0$ to $t=\infty$ is
then:
\begin{align*}
\Delta S_\text{sys}
  = \sum_\microst \left[ -\pi_\tau(\microst) \ln \pi_\tau(\microst)
  + \pi_0(\microst) \ln \pi_0(\microst) \right]
  ~.
\end{align*}
The trajectory-wise system entropy difference is defined to be:
\begin{align*}
\Delta s_\text{sys}(\traj)
= \ln \frac{\pi_0(\microst_0)}{\pi_\tau(\microst_\tau)}
 ~,
\end{align*}
where $\microst_0$ and $\microst_\tau$ are the initial and final microstates of
system microstate trajectory $\traj$, respectively. Averaged over all
trajectories $\traj \in \alltraj$, this then becomes the ensemble entropy
change.

Let $p(\traj | \microst_0)$ denote the probability of obtaining system
microstate trajectory $\traj$ via the protocol conditioned on starting the
system in state $\microst_0 = \traj(0)$.

Now, start the system Hamiltonian at $\mathcal H_\tau$ and run the reverse
protocol, ending the Hamiltonian at $\mathcal H_0$. We then obtain the
trajectory $\traj$ with a different conditional probability: $r(\traj |
\microst_0)$.

Assuming microscopic reversibility and given a system trajectory $\traj$, the
change in the heat bath's entropy is:
\begin{align}
\label{eq:MR}
\Delta S_\text{res}(\traj) = - \beta Q(\traj)
= \ln \frac{p(\traj | \microst_0)}{r(\rev \traj | \rev \microst_\tau)}
 ~,
\end{align}
where $\beta = 1 / \kB T$, $Q(\traj)$ is the net energy that flows out of the
heat bath into the system given the trajectory $\traj$, and $\rev{(\cdot)}$
denotes time-reversal. This holds for systems with strictly finite energies and
Markov dynamics that induce the equilibrium distribution when control
parameters are held fixed \cite{Croo99b}. Both our simulated Duffing potential
system and flux qubit obey these requirements at sufficiently short time
scales. Then we can express the total trajectory-wise change in entropy
production due to a trajectory $\traj$ as the sum of system and heat reservoir
entropy changes:
\begin{align*}
\Delta S_\text{tot}(\traj)
&= \Delta s_\text{sys}(\traj) + \Delta S_\text{res}(\traj)
 ~.
\end{align*}

Since $\pi_t(\microst) = e^{-\beta(\mathcal H_t (\microst) - \Feq_t)}$,
with $\Feq_t$ the system's equilibrium free energy at time $t$, we can write:
\begin{align*}
\Delta S_\text{tot}(\traj)
&= - \ln\pi_\tau(\microst_\tau) + \ln \pi_0(\microst_0)
- \beta Q(\traj) \\
&= \beta \left(\mathcal H_\tau(\microst_\tau) - \Feq_\tau \right)
- \beta \left(\mathcal H_0(\microst_0) - \Feq_0 \right) 
- \beta Q(\traj) \\
&= \beta \left(\Delta \mathcal H (\traj) - Q(\traj) - \Delta \Feq \right) \\
&= \beta \left(W(\traj) - \Delta \Feq \right)
 ~.
\end{align*}

Using Eq. (\ref{eq:MR}), we also have:
\begin{align*}
\Delta S_\text{tot}(\traj)
&= \Delta s_\text{sys}(\traj) + \Delta S_\text{res}(\traj) \\
&= \ln \frac{\pi_0(\microst_0)}{\pi_\tau(\microst_\tau)}
\frac{p(\traj | \microst_0)}{r(\rev \traj | \rev \microst_\tau)} \\
&= \ln \frac{\forP(\traj)}{\revPeq(\rev \traj)} \\
\intertext{with:}
\forP(\traj) &= \pi_0(\microst_0) p(\traj | \microst_0) \text{~and}\\
\revPeq(\rev \traj) &= \pi_\tau(\microst_\tau) r(\rev \traj | \rev{\microst_\tau})
 ~.
\end{align*}
Combining, we obtain a detailed fluctuation theorem:
\begin{align}
\label{eq:DFT}
\revPeq(\rev \traj)
&= \forP(\traj) e^{- \beta \left( W(\traj) - \Delta \Feq \right)}
 ~.
\end{align}

From here, we derive our first TCFT by integrating each side of Eq.
(\ref{eq:DFT}) over all trajectories $\traj$ in a measurable set $C \subset
\alltraj$. Starting with the LHS and recalling the Iverson bracket $[\cdot]$,
which is $1$ when the interior expression is true and $0$ when false, we have:
\begin{align*}
\int d\traj [ \traj \in C] \revPeq(\rev \traj)
& = \int d\rev \traj [ \traj \in C] \revPeq(\rev \traj) \\
& = \int d\rev \traj [ \rev \traj \in \rev C] \revPeq(\rev \traj) \\
& = \int d\traj [\traj \in \rev C] \revPeq(\traj) \\
& = \revPeq(\rev C)
 ~.
\end{align*}
The first three steps used the unity of the Jacobian in reversing a microstate,
the definition $\rev C = \{\rev \traj | \traj \in C\}$, and swapping all
instances of $\rev \traj$ with $\traj$. Integrating the RHS of Eq.
(\ref{eq:DFT}) then gives:
\begin{align*}
&\int d\traj [\traj \in C] \forP(\traj)
e^{- \beta \left( W(\traj) - \Delta \Feq \right)} \\
&\hspace{40pt}= e^{\beta \Delta \Feq} \int d\traj \forP(\traj, C)
e^{- \beta W(\traj)} \\
&\hspace{40pt}= \forP(C) e^{\beta \Delta \Feq} \int d\traj \forP(\traj | C)
e^{- \beta W(\traj)} \\
&\hspace{40pt}= \forP(C) e^{\beta \Delta \Feq}
\langle e^{- \beta W} \rangle_C
 ~.
\end{align*}
Combining, we have our first TCFT, Eq. (\ref{eq:tcft_expavg_work}).

To obtain the second TCFT, we first change the form of Eq. (\ref{eq:DFT}):
\begin{align*}
W(\traj)
  &= \Delta \Feq - \beta^{-1} \ln \frac{\revPeq(\rev \traj)}{\forP(\traj)}
 ~.
\end{align*}
Then we calculate the class-average. The equilibrium free energy
change is unaffected while the rightmost term becomes:
\begin{align*}
- \beta^{-1}
\left \langle \ln \frac{\revPeq(\rev \traj)}{\forP (\traj)} \right \rangle_C
&= - \beta^{-1} \int_C d\traj \forP(\traj | C)
\ln \frac{\revPeq(\rev\traj)}{\forP(\traj)} \\
&= - \beta^{-1} \int_C d\traj \forP(\traj | C)
\ln \frac{\revPeq(\rev\traj | \rev C) \revPeq(\rev C)}
{\forP(\traj | C) \forP(C)} \\
&= - \beta^{-1} \left( \int_C d\traj \forP(\traj | C)
\ln \frac{\revPeq(\rev\traj | \rev C)}{\forP(\traj | C)}
+ \ln \frac{\revPeq(\rev C)}{\forP(C)} \right) \\
&= \beta^{-1} \left(
D_\text{KL}\left[ \forP(\traj | C) || \revPeq(\rev \traj | \rev C) \right]
+ \ln \frac{\forP(C)}{\revPeq(\rev C)} \right)
 ~,
\end{align*}
which gives Eq. (\ref{eq:tcft_avgwork_equality})'s TCFT.

\section{Class-Averaged Work Approximation for Narrow Distributions}
\label{sec:app_expansion}

Here, we demonstrate that the class-averaged work $\langle W \rangle_C$
approaches its bound $\langle W \rangle_C^\text{min}$ when the variance and
higher moments of the class' distribution of works vanish. One concludes that
$\langle W \rangle_C^\text{min}$ is a good approximation for $\langle W
\rangle_C$ when the class' work distribution is narrow.

We first express the LHS of Eq. (\ref{eq:tcft_expavg_work}) in terms of the
unitless distance of work from its class-average:
\begin{align*}
\langle e^{-\beta W} \rangle_C
&= \langle e^{-x} \rangle_C \, e^{-\beta \langle W \rangle_C}
 ~,
\end{align*}
with $x = \beta (W - \langle W \rangle_C)$. Then, we Taylor expand the
exponential inside the class-average:
\begin{align*}
\langle e^{-x} \rangle_C
&= \sum_{n=0}^\infty \frac{(-1)^n}{n!} \langle x^n \rangle_C \\
&= 1 + a
 ~,
\end{align*}
with $a = \sum_{n=2}^\infty \frac{(-1)^n}{n!} \langle x^n \rangle_C$.
Equation (\ref{eq:tcft_expavg_work}) then gives:
\begin{align*}
(1 + a) e^{-\beta \langle W \rangle_C}
 & = \frac{\revPeq(\rev C)}{\forP(C)}e^{-\beta \Delta \Feq}
 ~.
\end{align*}
Since $e^{-x}$ is convex,
\begin{align*}
(1 + a) = \langle e^{-x} \rangle_C
\geq e^{-\langle x \rangle_C}
=1
 ~,
\end{align*}
so $a \geq 0$.
Then:
\begin{align*}
\langle W \rangle_C
&= \Delta \Feq + \beta^{-1} \ln \frac{\forP(C)}{\revPeq(\rev C)}
+\beta^{-1} \ln (1+a) \\
&\geq \Delta \Feq + \beta^{-1} \ln \frac{\forP(C)}{\revPeq(\rev C)} \\
&= \langle W \rangle_C^\text{min}
 ~.
\end{align*}
The second line becomes an equality when $a$ goes to zero, which occurs as the
variance and higher moments vanish.

\newcommand{\Hh}{\mathcal H}
\newcommand{\Hhs}{\mathcal H'}
\newcommand{\Hb}{H_\text{B}}
\newcommand{\Hs}{H_\text{S}}
\newcommand{\Hl}{H_\text{L}}
\newcommand{\Hi}{H_i}
\newcommand{\hsb}{h_\text{S,B}}
\newcommand{\hsl}{h_\text{S,L}}
\newcommand{\Hsb}{\mathcal H'}

\newcommand{\ddt}[1]{\frac{d#1}{dt}}
\newcommand{\pp}[2]{\frac{\partial #1}{\partial #2}}
\newcommand{\poi}[2]{\{#1, #2\}}

\newcommand{\Winc}{W}
\newcommand{\Wexc}{W_0}
\newcommand{\Qinc}{Q}
\newcommand{\Qexc}{Q_0}

\newcommand{\flux}{\phi}

\section{Work Definitions and Experimental Estimation}
\label{sec:ThermoWorks}


Properly estimating the required works and devolved heats from experimental
devices undergoing cyclic control protocols requires explicitly and
consistently accounting for energy and information flows between the system,
its environment, and the controlling laboratory apparatus. To this end, we
construct a model Hamiltonian universe for common processes involving small
systems interacting with laboratory apparatus and a thermal environment. After
deriving key equalities for two definitions of work, the inclusive and
exclusive works, we define a method of approximating them in appropriate cyclic
protocols.


\subsection{The Model Universe and Hamiltonian}

To study a small system that exchanges energy with its environment in the forms
of heat and work, we introduce a model universe: a \emph{system of interest}, a
\emph{heat bath}, and a \emph{lab} (laboratory apparatus) that controls the
system and derives any needed energy from a \emph{work reservoir}. The system
directly interacts with both the heat bath and the lab, but the heat bath and
lab are not directly coupled.

We assume that a Hamiltonian $\Hh$  describes the universe's evolution and that
there is a set of generalized coordinates which can be sensibly partitioned
into those for the system, heat bath, and lab.  Then, we decompose the universe
Hamiltonian into the following form:
\begin{align*}
\Hh(s, b, l) = \Hb(b) & + \hsb(s, b) + \Hs(s) + \hsl(s, l) + \Hl(l)
 ~,
\end{align*}
where $s$, $b$, and $l$ denote both the generalized coordinates and conjugate
momenta for the system, bath, and lab, respectively. For any universe
Hamiltonian $\Hh$, there can be many choices for this decomposition.

We also define the system Hamiltonian $\Hhs$ as the three components
that depend on the system coordinates:
\begin{align*}
\Hhs(s; b, l)
= \hsb(s, b) + \Hs(s) + \hsl(s, l)
 ~.
\end{align*}

First, consider the subset of lab coordinates $l$ for which $\hsl$ has
nontrivial dependence. These so-called \emph{protocol parameters} $\lambda$ are
often simple and much fewer than the entire set of $l$. We often assume that we
have total control of their evolution. More precisely, under an appropriate
preparation for the lab at time $t=0$, a specific trajectory for the protocol
parameters $\{\lambda(t)\}_t$ for $0 \leq t \leq \tau$ is guaranteed for all
preparations of the heat bath and system coordinates. We refer to the parameter
trajectory as the \emph{protocol}.

Suppose the heat-bath degrees of freedom that interact with the system change
much faster than the system's. We can assume that the system response to the
bath resembles Brownian motion. On the time scale of changes in the system
coordinates, then, we ignore the system-bath interaction term $\hsb$ in writing
the system Hamiltonian $\Hhs$:
\begin{align*}
\Hhs(s; \lambda) & = \Hs(s) + \hsl (s, \lambda) \\
  & = T(s) + V(s, \lambda)
  ~.
\end{align*}
The latter decomposition into kinetic energy $T$ and potential energy $V$ can
be used to write Langevin equations of motion for the system. Furthermore, if
the heat bath has a relaxation time sufficiently short that it is roughly in
equilibrium at all times with fixed temperature, then its influence on the
system will be memoryless.

\subsection{Inclusive and Exclusive Works and Heats}

The basic scenario for executing a protocol is as follows. The universe
coordinates begin according to a given initial distribution $\Pr(s)$ at time
$t=0$ and they evolve in isolation until $t=\tau$. As above, we assume that a
well-defined protocol $\{\lambda_t\}_t$ emerges due to our preparation of the
lab coordinates.

We label all energy exchanged between the system and lab as \emph{work} and all
energy exchanged between the system and heat bath as \emph{heat}. Since the lab
is directly coupled only to the system, the work they exchange is given by the
change in energy of the lab's work reservoir. Similarly, since the heat bath is
directly coupled only to the system, the heat exchanged is given by the change
in the heat bath's energy.

Note that this requires choices as to what constitute the energies of the three
universe subsystems. While $\Hb$, $\Hs$, and $\Hl$ define energies for the heat
bath, system, and work reservoir, respectively, what of $\hsl$ and $\hsb$?  If
all subsystems were macroscopic, these interaction terms would be negligible.
While it may be desirable to assume that the system is only weakly coupled to
the heat bath---so that $\hsb$ can be ignored---$\hsl$ can be significant in
many important small systems.

And so, in general, we define the system energy to be $\Hs$ plus any portions
of $\hsl$ and $\hsb$. Then the work reservoir energy is $\Hl$ plus the rest of
$\hsl$, while the heat bath energy is $\Hb$ plus the rest of $\hsb$. To make
these distinctions clear we label two types of works, each corresponding to
the two extremes for allocation of $\hsl$ between the system and work
reservoir: the \emph{inclusive work} $\Winc$ and the \emph{exclusive work}
$\Wexc$
\cite{Jarz07a}.
Specifically:
\begin{alignat*}{2}
\ddt \Winc &= - \ddt{} (\Hl) &&\,= \ddt{} (\Hb + \hsb + \Hs + \hsl) \\
\ddt \Wexc &= - \ddt{} (\hsl + \Hl) &&\,= \ddt{} (\Hb + \hsb + \Hs)
 ~.
\end{alignat*}
We can similarly define the \emph{inclusive heat} $\Qinc$ and \emph{exclusive
heat} $\Qexc$
depending on how we allocate $\hsb$ between the system and heat bath:
\begin{alignat*}{2}
\ddt \Qinc &= -\ddt{} (\Hb) &&\,= \ddt{} (\hsb+ \Hs + \hsl + \Hl) \\
\ddt \Qexc &= -\ddt{} (\Hb + \hsb) &&\,= \ddt{} (\Hs + \hsl + \Hl)
 ~.
\end{alignat*}
The inclusive work corresponds to fully including $\hsl$ in the system
energy, while the exclusive work corresponds to excluding it.
Inclusive and exclusive heat correspond similarly with respect to $\hsb$.

There is a key relation between the inclusive and exclusive works:
\begin{align}
\ddt \Winc &= \ddt \Wexc + \ddt \hsl
 ~.
\label{eq:W_inc_exc_relation}
\end{align}
That is, the inclusive work for an interval of time equals the sum of the
exclusive work and the change in the system-lab interaction term $\hsl$.

In the above expressions, calculating the rate of change of a work or heat
requires the time derivative of one or more of $\Hl$ and $\Hb$. This can be
problematic. Fortunately, there are alternate forms that are amenable.
One can show that the inclusive work rate is given by:
\begin{align}
\ddt \Winc &= -\ddt \Hl \nonumber \\
  & = \pp \hsl \lambda \ddt \lambda
\label{eq:inc_work_eq}
 ~.
\end{align}
This is a more common definition for the work rate in small-system
nonequilibrium thermodynamics. And, it allows the work to be calculated as:
\begin{align}
W(t, t') = \int_{t}^{t'} dt'' \frac{d\lambda}{dt''}
\pp {\hsl(s, \lambda)}{\lambda} |_{\lambda = \lambda(t'')}
 ~.
\end{align}
The exclusive work $W_0$ has a corresponding form:
\begin{align}
\ddt \Wexc & = -\ddt {(\hsl + \Hl)} \nonumber \\
  & = - \pp \hsl s \ddt s
\label{eq:exc_work_eq}
 ~,
\end{align}
For the case where $\hsl$ is a scalar potential for $s$, this is the product
of the corresponding force with velocity. This makes the exclusive work
equal to a familiar mechanics definition of work as the integral of the dot 
product of force and displacement:
\begin{align*}
\Wexc(t, t') = - \int_t^{t'} dt'' \frac{ds}{dt''}
\pp \hsl s |_{s=s(t'')}
 ~.
\end{align*}
In this way, we write the inclusive and exclusive work rates in terms of the
rates of change of the system and work-reservoir interaction term $\hsl$
with respect to either the system or work reservoir coordinates.

\subsection{Approximating Inclusive Work Experimentally}

For the flux qubit experimental system investigated here, we assume the
following:
\begin{align*}
\Hs(s) + \hsl(s, \lambda) = T(s) + V(s, \lambda)
 ~.
\end{align*}
That is, as far as the flux qubit and work reservoir are concerned, the only
relevant energies at least partially ascribable to the flux qubit are its
kinetic energy and the potential energy with the work reservoir.
$\hsl$ must then capture the change in the potential $V$ due to changes in the
protocol parameters. We could simply define $\hsl(s, \lambda) = V(s, \lambda)$
so that $\Hs(s) = T(s)$. However, it is more useful to allocate the initial
potential energy to $\Hs$. That is:
\begin{align*}
\Hs(s) &= T(s) + V(s, \lambda_0) ~\text{and}\\
\hsl(s, \lambda) &= V(s, \lambda) - V(s, \lambda_0)
 ~.
\end{align*}
For cyclic protocols where $V(\cdot, \lambda_0) = V(\cdot, \lambda_\tau)$, such
as in our erasure operation, $\hsl(s(t), \lambda(t))$
vanishes for all trajectories at $t=0, \tau$. By Eq.
(\ref{eq:W_inc_exc_relation}) we then have the useful equality $\Winc = \Wexc$
between inclusive and exclusive works taken over the entire protocol.

Estimating $\Winc$ for a system trajectory is then equivalent to estimating
$\Wexc$ for the cyclic protocols we consider. In the flux qubit, the form of
$\hsl$ is known and the specific protocol $\{\lambda_t\}_{t \in [0,\tau]}$ is
known. Unfortunately, we lack sufficient information about its instantaneous
state $s$ at all times, since the device's physics precludes precise
measurements of system flux $\flux$---the relevant part of $s$ for determining
the potential $\hsl$. Instead, we do have reliable measurement of large and
stable changes in the flux $\flux$. This specifically monitors when the system
moves between wells in a double-well potential $V(\cdot, \lambda(t))$, if the
rate of transition between wells is sufficiently slow.

And so, we can use information about the flux $\flux$ to approximate the
exclusive work contribution at each moment in time. Then, adding up these
contributions yields an approximation to the total exclusive work $\Wexc$ over
the entire protocol and therefore of the inclusive work $\Winc$ over the entire
protocol. Note that the protocols used here maintain two wells at all times for
the system flux $\flux$. We develop the approximation in two steps.

\subsubsection{First-Order Approximation}

We first partition the potential in flux space into three segments. Two segments
constitute the wells for the flux in which that state spends all its time
except for very brief transitions between wells. Then, the third segment
connects the two wells, capturing the dynamics arising from crossing the
barrier that separates them.

We require that the partitioning allows the following two approximations.
First, the particle spends negligible total duration in between the two wells.
Second, the wells do not change shape over the protocol, but instead simply
raise or lower in potential at different times, if they change at all. This
means that the shape of the system-lab interaction term $\hsl(\cdot,
\lambda(t))$ at any time $t$ is very simple in the two wells---flat.

The result is that the exclusive work over any time duration is easily
calculable from the experimental data. During times when the flux remains in a
well, the exclusive work must be zero, since $\hsl$ does not change with $s$.
During a transition, the shape of $\hsl$ does not change due to the first
approximation. Then, the exclusive work is the difference in heights of the two
wells as measured by $\hsl$:
\begin{align}
\Delta \Wexc^\text{trans} &= \int d\flux \left( -\pp \hsl \flux \right) \nonumber \\
&\approx \hsl(w_0, \lambda(t)) - \hsl(w_1, \lambda(t))
  ~,
\label{eq:jump_contribution}
\end{align}
where $\lambda(t)$ is the protocol parameter setting at any time during the
transition and $w_0$ and $w_1$ are arbitrary flux values in the starting and
ending wells, respectively.

Thus, the total inclusive work $W$ over the protocol for a trajectory is simply
the sum of the jump contributions above for each transition.

\subsubsection{Second-Order Approximation}

In point of fact, the potential wells do change shape. Fortunately, our method
for calculating the inclusive work over the protocol remains valid under weaker
constraints on the protocol.

We first require the protocol to maintain two metastable regions, the
\emph{informational states}, at all times; each possessing a unique local
potential minimum continuously in time. We denote the flux value at the
potential minima of informational state $i$ at time $t$ as $\flux_i^t$. The
protocol must also evolve slowly enough so that the potential landscape changes
slowly compared to the system's relaxation rate in each metastable region.
Both of these criteria are met by our erasure protocol.

Consider a short duration $\Delta t$ during which the potential $V(\cdot, t)$
changes little but long enough compared to the relaxation rates of the
informational states. Consider two cases: either the system crosses the barrier
between the two informational states during this time or it remains in one
informational state.

First, suppose that the system transitions from one informational state $i$ to
the other $j$. Denote the system flux at the beginning of the transition as
$\flux^t$ and at the end as $\flux^{t + \Delta t}$. By Eq.
(\ref{eq:W_inc_exc_relation}), the exclusive work contribution $\Delta
\Wexc^\text{trans}$ is the difference of the inclusive work contribution and
the change in system-lab interaction term $\Delta \hsl$. The change $\Delta
\hsl$ can itself be broken down into two terms, one for the difference in
$\hsl$ between the informational-state minima and the other for the change in
$\hsl$ local to the respective minima. In other words:
\begin{align}
\Delta \hsl & = \hsl(\flux^{t + \Delta t}) - \hsl(\flux^t) \nonumber \\
  & = \big[ \big(\hsl (\flux^{t + \Delta t}) - \hsl(\flux_j^{t+\Delta t})\big)
+ \hsl(\flux_j^{t+\Delta t}) \big]
  - \big[ \big(\hsl (\flux^t) - \hsl(\flux_i^t)\big)
  + \hsl(\flux_i^t) \big] \nonumber \\
  & = \big[ \hsl(\flux_j^{t+\Delta t}) - \hsl(\flux_i^t) \big]
  + \big[ \big(\hsl (\flux^{t + \Delta t})
  - \hsl(\flux_j^{t+\Delta t})\big)
   - \big(\hsl (\flux^t) - \hsl(\flux_i^t)\big) \big]
  \label{eq:hsl_prebreakdown} \\
   & = \Delta m^t + \Delta l^t
\label{eq:hsl_breakdown}
 ~,
\end{align}
where $\Delta m^t$, Eq. (\ref{eq:hsl_prebreakdown})'s first term, is the change
in $\hsl$ at the informational-state minima and $\Delta l^t$, Eq.
(\ref{eq:hsl_prebreakdown})'s second term is the change in $\hsl$ of the system
with respect to the informational-state minima. Our protocol ensures that the
total number of transitions is so small and the time durations so narrow that
we can ignore the total contributions of inclusive works $\Delta
\Winc^{\text{trans}}$ due to these transition durations. Then, we approximate
the exclusive work contribution during a transition via:
\begin{align*}
\Delta \Wexc^\text{trans} &= -\Delta \hsl \\
  & = - \Delta m^t - \Delta l^t
 ~.
\end{align*}

Suppose, now, that the system remains in one informational state $i$ during a
time interval $\Delta t$. Since the relaxation rate is fast compared to the
duration $\Delta t$, we assume that the system visits all microstates in the
informational state roughly in proportion to the local equilibrium
distribution. Then, the inclusive work contribution $\Delta \Winc^\text{stay}$
is approximately independent of the specific system trajectory during this time
and, instead, is determined by the time duration and the informational state
$i$. If during this time we simultaneously shift the entire potential up by a
given amount, we add an inclusive work contribution equal to the potential
shift but the system trajectory is unchanged. Thus, the actual inclusive work
contribution is equal to an amount due to the change in the system-lab
interaction term at the informational-state minimum plus an amount due solely
to the change in potential shape at the informational state with respect to its
minimum. That is:
\begin{align}
\Delta \Winc^\text{stay} = \Delta \Winc_s + \Delta m^t
 ~,
\end{align}
where $\Delta \Winc_s$ is the inclusive work contribution due to the change in
potential shape at the informational state. Equation (\ref{eq:hsl_breakdown})
applies equally well here in describing the change in system-lab interaction
term. Thus, the exclusive work contribution $\Delta \Wexc^\text{stay}$ for this
time interval is:
\begin{align}
\Delta \Wexc^\text{stay} &= \Delta \Winc^\text{stay} - \Delta \hsl \\
&= \Delta \Winc_s - \Delta l^t
 ~.
\end{align}
The result is that we have exclusive work contributions for both durations when
the system transitions between informational states and when it remains in one.

To find the total exclusive work over the protocol for a given trajectory we
add up the contributions. The sum of all local $\hsl$ changes $\Delta l^t$ over
all durations is the net local change in $\hsl$. Recall, though, that the
minima of the informational states begin and end at the same values. And so,
the total local change in $\hsl$ reduces to the absolute change in $\hsl$.
However, since we chose $\hsl(\cdot, 0) = \hsl(\cdot, \lambda(t)) = 0$, this
must vanish:
\begin{align*}
\sum_t \Delta l^t = 0
 ~.
\end{align*}

We can now specify our final approximation: At any time $t$, the inclusive work
contribution $\Delta W_s$ due to the change in potential shape is independent
of the informational state. This is reasonable for our erasure protocol since
the asymmetric contribution to the change in potential---the tilt---is slight.
While it clearly breaks the symmetry of the double-well potential by changing
the well heights, it has less effect on the well shapes and even less in making
those shapes distinct.

Then, we can assume that the sum of $\Delta W_s$ for a trajectory is the same
as that staying in one informational state the entire time. Since the protocol
is very slow and cyclic, though, a particle that stays in one informational
state the entire time must receive approximately zero inclusive work $\Winc$.
Given that the sum over all $\Delta W_s$ must be equal to $\Winc$ for such a
trajectory, it must also be negligible.

Altogether, the total exclusive work is approximately given by the sum over all
transitions between informational states of the difference in potential at the
informational-state minima:
\begin{align}
\Wexc(0, \tau) = - \sum_\text{trans} \Delta m^t
 ~.
\end{align}
To reiterate, since $\Delta \hsl = 0$, this is also the total inclusive work
$\Winc(0, \tau)$ for a trajectory over the entire protocol.

\section{Substage Work Distributions Commentary}
\label{sec:SubstageWorkDists}

Here, we briefly interpret several features of the substage work distributions
observed in Fig. \ref{fig:protocol}(Outer left plots).

The distributions for barrier dropping and tilting are narrow, symmetric peaks;
see Fig. \ref{fig:protocol} (Outer left plots).
Barrier raising also has a rather narrow peak, composed primarily of
trajectories always in the $\RightISt$ state, but also exhibits a bulge toward
positive work; see Fig. \ref{fig:protocol} (Top right). Note that the
$\LeftISt$ state is created mid-way through barrier raising, allowing for
trajectories that spend some time in either informational state, but
disallowing trajectories that spend all time in the $\LeftISt$ state. The
former induce the positive work bulge toward less negative works, which while
notable will not be further explored here.

The substage work distributions for untilting presents the most striking
picture; see Fig. \ref{fig:protocol} (Bottom right).
Always-$\RightISt$ trajectories induce a large positive work peak (red),
always-$\LeftISt$ trajectories induce a large negative work peak (orange), and
all other trajectories induce a ramp between them (blue).

These features can be directly interpreted by following the locations of the
potential minima over time and noting how the shifting potential adds or
removes energy from a particle. During barrier dropping, to take one example,
the protocol raises both minima by over $7~\kB T$, resulting in a
narrow, peaked work distribution with a mean near $7~\kB T$.

Most interesting is the untilt substage. Since most particles start and then
stay in the $\RightISt$ state for this substage, a large positive work is
probable, due to the rising $\RightISt$-state well. However, it is also
possible for the system to start in and then get stuck in the $\LeftISt$-state
well, resulting in a large negative work. The final possibility is
transitioning between states during untilting, resulting in an intermediate
range of less-likely work values. For trajectories that do transition between
states during untilting, it is more likely to spend more time in the
$\RightISt$ state, since it is energetically favored, resulting in the rising
probability with increasing work in their work distribution---giving rise to
the log-linear ramp in the work distribution.

Note that there are small peaks on each end of this third class' distributions
that require a more nuanced explanation. When a particle crosses a
barrier---due to random thermal excitation---the surplus energy may quickly
send the particle back to the previous well before it can be dissipated. Such
particles then spend almost all of the substage in this first well, generating
a work value accordingly. Statistics of the ramp proper are due to particles
that have time to locally equilibrate before crossing any barriers.

Follow-on work develops the theory underlying this detailed mechanistic
analysis and analyzes similar behavior in all metastable-quasistatic processes.

\section{Flux Qubit Device, Calibration, and Measurement}
\label{sec:FluxQubit}

The benefits of the flux qubit device are several-fold. First, their physics
provide a genuine two-degree of freedom dynamics, while other comparable
experiments on Maxwellian demons and bit erasure are very high dimensional,
only indirectly providing an effectively few-degree of freedom dynamics
\cite{Beru12a,Beru13a,Jun14a}. Second, they operate at very high frequency and
so one readily captures the substantial amounts of data required to accurately
estimate rare-fluctuation statistics. Third, they leverage recent advances in
manufacturing technology led by efforts in quantum computing. Fourth, being
constructed via  modern integrated circuit technology they form the basis of a
technology that will scale to large, multicomponent circuit devices for more
sophisticated thermodynamic computing. And, finally, in the near future flux
qubits will facilitate experiments that probe the thermodynamics of the
transition to quantum information processing.

At the microscopic level, a fraction of the electrons in a superconducting
metal form bosonic Cooper pairs---a quantum-coherent condensate. For designing
superconducting electronic circuits, though, one can forgo the microscopic
description and work with higher-level phenomena, such as flux quantization and
the Josephson relations for weak links. Importantly, the circuit-level degrees
of freedom are not coarse-grained quantities, but display a full range of
quantum behavior, including quantized excitations, coherent superpositions, and
entangled states in such circuits. For our purposes here, however, we run the
device so that it exhibits only classical stochastic dynamics, reserving
quantum information thermodynamic explorations for the future.

This section lays out the basic physics of the flux qubit device and details of
the experimental implementation.

\subsection{Flux qubit physics}
\label{sec:FluxQubitPhysics}

Our experimental information processor is a special type of superconducting
quantum interference device (SQUID) with two degrees of freedom---a
gradiometric flux qubit or the variable-$I_c$ rf SQUID introduced by Ref.
\cite{Han92a}.
Notably, the energies associated with the motion
perpendicular to and along the escape direction differ substantially by about a
factor of $12$. Practically, this asymmetry reduces the two-dimensional
potential to one dimension. The net result is a device with an effective
double-well potential with barriers as low as $\Delta U \sim \kB T$ that
operates at frequencies in the GHz range. The potential shape is controlled
by fluxes that are readily controlled by currents. SQUID device parameters,
used to determine the potential shape and energy scales, were all independently
determined.

The variable-$I_c$ rf SQUID replaces the single Josephson junction in a standard
rf SQUID with a symmetric dc SQUID with small inductance $\beta_\text{dc} = 2
\pi \ell I_{c0}/\Phi_0 \ll 1$, where $2 \ell$ is the loop inductance, $I_{c0} =
i_{c1} + I_{c2}$ is the sum of critical currents of the two junctions, and
$\Phi_0$ is the flux quantum $h/2e$. This architecture gives a device whose
parameters can be accurately measured and that can be selected to exhibit a
range of phenomena including thermal activation, macroscopic quantum tunneling,
incoherent relaxation, photon-induced transitions, and macroscopic quantum
coherence. It also allows us to perform, as we demonstrate, nanoscale
thermodynamic computing.

Its macroscopic dynamical variables are the magnetic flux $\Phi$ through the
rf SQUID loop and $\Phi_\text{dc}$ through the dc SQUID loop. Based on the
resistively-capacitively-shunted junction model of Josephson junctions, in the
classical limit the variable-$I_c$ rf SQUID's deterministic equations of motion are
\cite{Han92a}:
\begin{align}
2C \ddot{\Phi} + \frac{\dot{\Phi}}{R/2}
  & = -\frac{\partial U(\Phi,\Phi_\text{dc})}{\partial \Phi} ~\text{~and}
  \nonumber \\
  \frac{C}{2} \ddot{\Phi}_\text{dc} + \frac{\dot{\Phi}}{2R}
  & = -\frac{\partial U(\Phi,\Phi_\text{dc})}{\partial \Phi_\text{dc}}
  ~.
\label{eq:FluxQubitEoM}
\end{align}
In units of $\Phi_0/2 \pi$, the 2D potential for the variable-$I_c$ rf
SQUID is $U(\phi,\phi_\text{dc}) = U_0 f(\phi,\phi_\text{dc})$ with: 
\begin{align}
f(\phi,\phi_\text{dc}) = \tfrac{1}{2} (\phi-\phi_x)^2 + \tfrac{\gamma}{2}
(\phi_\text{dc} - \phi_\text{xdc})^2
- \beta_0 \cos \tfrac{\phi_\text{dc}}{2} \cos \phi
+ \delta \beta \sin \tfrac{\phi_\text{dc}}{2} \sin \phi
  ~,
\label{eq:FluxQubitPotential}
\end{align}
where $U_0 = \Phi^2_0/ (4 \pi^2 L)$. Here, $\gamma = L/(2 \ell)$ is the ratio of
rf and dc SQUID inductances; $\phi_x$ ($\phi_\text{xdc}$) is the external
flux applied to the rf (dc) SQUID loop; $\phi$ ($\phi_\text{dc}$) is the flux
enclosed in the rf (dc) SQUID loop; $\beta_0 = 2 \pi L I_{c0}/\Phi_0$; and
$\delta \beta = 2 \pi L (I_{c2}-I_{c1})/\Phi_0$.

For large-amplitude tuning of the external controls, the system response to
$\phi_x$ ($\phi_\text{xdc}$) is $2\pi$ ($4\pi$) periodic. We make use of the
global features to accurately determine the coefficients of the potential.

In the experiment, cross-coupling between the barrier and tilt controls was
canceled by an affine transformation
$(\phi_\text{x}, \phi_\text{xdc}) \rightarrow
(\phi_\text{x} + \alpha \phi_\text{xdc}, \phi_\text{xdc})$,
with the coefficient $\alpha$ chosen such that the equilibrium population of
the left and right wells was unaffected to first order by the barrier control
$\phi_\text{xdc}$.

Operating the magnetometer generates wide-band local electromagnetic
interference that can affect the dynamics of the flux qubit. A careful study of
the back-action indicates that low-amplitude operation of the magnetometer can
induce transitions in a manner that corresponds to a shift in the effective
tilt and flux controls. Importantly, the effective temperature under
magnetometer operations was not elevated from $500$ mK.

The dynamical variable $\phi$ describes the in-phase motion of the two
junctions that results in a current circulating in the rf SQUID loop. The
dynamical variable $\phi_\text{dc}$ describes the out-of-phase motion,
resulting in a current circulating in the dc SQUID loop. The shape of the
effective potential is completely determined by the dimensionless function
$f(\phi,\phi_\text{dc})$ and the energy scale of the potential is determined by
$U_0$. With suitable device parameters and applied fluxes ($\phi_\text{x}$ and
$\phi_\text{xdc}$) one obtains a double-well potential. The barrier height
$\Delta U$ separating the two wells is readily adjusted by varying
$\phi_\text{xdc}$. The effective potential is plotted in Fig.
\ref{fig:FluxQubit}(B) with parameters: $\beta_0 = 6.2$, $\gamma = 12$ and
$\delta_b = 0.2$.

\subsection{Experimental implementation}

The junctions were $1 \times 1 \mu \text{m}^2 \text{Nb/Al}_2
\text{O}_3/\text{Nb}$ tunnel junctions of very low subgap leakage, typically
having a quality factor of $V_m \approx 70$ mV at $4.2$ K.

We followed a standard procedure (see, e.g., Ref.
\cite{Han92a})
for
calibrating the flux qubit parameters. An outline of the steps is given below.
A complete description of the measurements is presented elsewhere.

First, by executing wide-range sweeps of the coil currents $I_\mathrm{tilt}$
and $I_\mathrm{barrier}$, parameter values corresponding to single-valued and
bistable potential landscapes are recorded. A linear transformation from
$I_\mathrm{tilt}$ and $I_\mathrm{barrier}$ to $(\phi_\text{x}, \phi_\text{xdc})$ is
established by matching the experimental periodicity with the theoretical one
$(2\pi,4\pi)$. Linear cross-talk from $I_\mathrm{barrier}$ to $I_\mathrm{tilt}$
is calibrated by orthogonalizing the global response. Cross-talk from
$I_\mathrm{tilt}$ to $I_\mathrm{barrier}$ can be assumed to be small due to the
symmetry of the on-chip flux lines and is taken to be zero.

The parameter values $\beta_0 = 6.2$, $\gamma = 12$ and $\delta_b = 0.2$ are
determined by equating the observed extent of hysteresis at $\phi_\text{xdc} =
0$ and the differential flux response $d\left<\phi\right> / d\phi_\text{x}$ at
$\phi_\text{xdc} = 2\pi$ to theoretical predictions. The prefactor $U_0 =
56.3$~K is determined by equating the observed escape energy for inter-well
transitions at high temperatures with $\kB T$. The plasma frequency $\omega_p =
1/\sqrt{LC} = 2\pi \times 13.7$~GHz is determined from the observed
low-temperature cross-over temperature $T_cr = 103$~mK to macroscopic quantum
tunneling (MQT) dominated dynamics. We obtain an upper bound $Q = \omega_p R C
< 130$ from the coupling to the passive shunt resistor of the magnetometer.
Parameter calibration measurements are performed in such a way that the effect
of magnetometer back-action is nulled through pulsing of the readout or
otherwise minimized. The effective temperature under continuous magnetometer
operation was determined by repeating the measurement for escape energy for
interwell transitions and comparing the result to that obtained under pulsed
magnetometer operation.


\end{document}